\documentclass[3p,twocolumn,10pt,nofootinbib,showkeys,showpacs,a4paper]{revtex4-1}

\newcommand{\versionnumber}{v3.00}

\renewcommand{\today}{KA--TP--35--2011 (\versionnumber)}
\usepackage{hyperref}
\usepackage{graphicx}
\usepackage{subfigure}
\usepackage{bm}
\usepackage{mathrsfs}
\usepackage[latin1]{inputenc}
\usepackage{stmaryrd}
\usepackage{array}
\usepackage{psfrag}
\usepackage{dsfont}
\usepackage{epsfig}
\usepackage{titletoc}
\usepackage{float}   %  for figures
\usepackage{wrapfig} %  for figures
\usepackage{amsmath}
\usepackage[latin1]{inputenc}
\usepackage{amsmath}\usepackage{amssymb,stmaryrd}
\usepackage{mathrsfs}
\usepackage{array}
\usepackage{graphicx}
\usepackage{psfrag}
\usepackage{dsfont}
\usepackage{epsfig}
\usepackage{cancel}
\usepackage{upgreek}
\usepackage{multirow}

\begin{document}

%\begin{frontmatter}

\title{Multiple Lorentz groups --- a toy model \\ for superluminal muon neutrinos}

\author{M. Schreck} \email{marco.schreck@kit.edu}
\affiliation{Institute for Theoretical Physics, Karlsruhe Institute of Technology (KIT),\\
         76128 Karlsruhe, Germany}

\begin{abstract}
In this article an idea is presented, which allows for the explanation of superluminal muon neutrinos. It is based on
the introduction of a new superluminal, massless gauge boson coupling to the neutrino only, but not to other standard
model particles.

The model is discussed with regard to the Supernova 1987 (SN 1987) velocity bound on electron antineutrinos and the
Cohen--Glashow constraint on superluminal neutrino propagation. The latter can be circumvented if --- within the
framework of the model --- a sterile neutrino mixing with the active neutrino mass eigenstates is introduced. The
suggestion of a sterile neutrino accounting for superluminal neutrinos has already been proposed in several papers.

It is possible to choose mixing angles with the sterile neutrino sector such that the model respects both the SN 1987
bound and the muon neutrino travels superluminally.
\end{abstract}

\maketitle
%\twocolumn

\setcounter{equation}{0}
\renewcommand{\theequation}{\arabic{section}.\arabic{equation}}
\section{Introduction}
\label{sec:Introduction}

At the end of September 2011, the OPERA collaboration announced the remarkable detection of superluminal muon neutrinos at
the $6.2\sigma$-level \cite{OPERA:2011zb}. Their experimental result with statistical and systematical error was given by:\footnote{
These numbers can be found in the updated version 2.0 of \cite{OPERA:2011zb}.}
\begin{equation}
\label{eq:opera-result-neutrino-velocity}
\frac{(v_{\upnu_{\mu}}|_{E_{\upnu_{\scalebox{0.4}{$\mu$}}}=17\,\mathrm{GeV}})-c}{c}=(2.37\pm 0.32\,{}^{+0.34}_{-0.24})\cdot 10^{-5}\,.
\end{equation}
Unfortunately, on February, 2012 two error sources had become evident, which were likely to ruin their result. First, a fiber
connection to a computer card had not been attached properly. Second, there had been a problem with the clock at OPERA used
between the synchronizations with the Global Positioning System.

At the 25th International Conference on Neutrino Physics and Astrophysics in Kyoto on June 8th, 2012 a final update on the OPERA
time-of-flight measurement was given:
\begin{equation}
\delta t=1.6\pm 1.1^{+6.1}_{-3.7}\,\mathrm{ns}\,.
\end{equation}
This number states the deviation of the muon neutrino time-of-flight from the time that light needs to travel the distance from
CERN to the Gran Sasso underground laboratory. Hence, the deviation is now consistent with zero.

The physics community had considered the result given by Eq.~\eqref{eq:opera-result-neutrino-velocity} with care, since the
deviation from the speed of light lay several orders of magnitude above what would be expected, if it was from quantum gravitational
origin. The authors of \cite{Contaldi:2011,Besida:2011fi,vanElburg:2011ze,Alicki:2011qw,Henri:2011wu,Broda:2011du} tried to
figure out how the OPERA result could be explained by possible systematical errors. Beyond that, in \cite{Davoudiasl:2011gb} a
cross-check for the result was proposed. It was demonstrated that muon neutrinos traveling with superluminal velocity can produce
signatures for highly-boosted $\mathrm{t\overline{t}}$-quark pairs at the LHC, where one or both quarks decay semileptonically.

Furthermore, on the one hand, Cohen and Glashow showed that a superluminal neutrino would lose its energy quickly by the emission
of electron positron pairs \cite{Cohen:2011hx}. If muon neutrinos moved faster than light, the process $\upnu_{\mu}\rightarrow \upnu_{\mu}\mathrm{e^+e^-}$
would be energetically possible above a certain neutrino energy threshold resulting in a copious production of electron positron
pairs. On the other hand, the authors of \cite{Mohanty:2011,Lingli:2011kh} discussed that the Cohen--Glashow constraint can be avoided.
This may be the case when, for example, Lorentz-violating effects depend quadratically on the neutrino energy or if Lorentz-violation
is not fixed but covariant with the neutrino four-momentum.

In \cite{arXiv:1111.4994} two models were investigated, where the first gave rise to deformed energy conservation laws and the
second resulted in deformed momentum conservation laws. For these models the bounds of \cite{Cohen:2011hx} are not applicable.

In the article \cite{Bi:2011nd} further constraints on the deviation of the neutrino velocity from the speed of light were given by
considering pion decay and TeV-neutrinos detected by ICECUBE. These gave severe bounds on Lorentz symmetry violation in the neutrino
sector clashing with the experimental OPERA result.

Although the OPERA result has now proven to be wrong, it stimulated theoretical ideas in this field and led to many new models, which
can perhaps be applied to other realms of physics. It may also be the case that a certain neutrino species indeed travels superluminally.
However, the deviation from the speed of light is then expected to be much smaller than the value measured by OPERA. For this reason
some representative examples for models that try to describe superluminal neutrinos will be listed:

\begin{itemize}

\item It is well-known that dispersion relations of particles will be modified, if they propagate through a medium. In
\cite{Cacciapaglia:2011ax,AmelinoCamelia:2011dx} the superluminal motion of muon neutrinos is interpreted in the framework of
deformed dispersion relations, which are a low-energy manifestation of Lorentz-violating physics at the Planck scale. In these theories
the vacuum behaves as an effective medium.

Such a medium can result from standard model physics, as well. For example, in \cite{Nakanishi:2011vv} superluminal neutrinos are
explained by the assumption that Earth is surrounded by a special kind of matter consisting of separated quarks. If the wave functions
of quarks are entangled, they can form colorless objects and, hence, are confined, even when they are spatially separated by a large
distance.
\item In \cite{Nojiri:2011ju} a model is proposed which describes a spontaneous breakdown of Lorentz symmetry by a scalar background
field that is added to the action via a Lagrange multiplier. This framework leads to a modified neutrino dispersion relation depending
on the momentum of the neutrino.
\item The neutrino velocity can be modified by Fermi point splitting (for a recent review see \cite{arXiv:1111.4627}),
which removes the degeneracy of zeros of the fermionic energy spectrum \cite{Klinkhamer:2011mf}.
\item The neutrino dispersion relation can change because of environmental effects caused by fields that accumulate at the position
of the Earth. These may lead to an effective metric, in which the neutrino propagates with superluminal velocity
\cite{Dvali:2011mn,Oda:2011kh}.\footnote{Reference~\cite{Iorio:2011ay} gives new experimental bounds on the mass scale $M_{*}$
that is characteristic for the model presented in \cite{Dvali:2011mn}.} Furthermore, in the context of general relativity, a
particle traveling along a geodesic path in a metric different from the Minkowski metric can be investigated \cite{Alles:2011wq}.
In the article previously mentioned the mean velocity of such a particle is calculated with the assumption that the observer stays at
rest. The average velocity can be larger than the speed of light, even if the velocity as a local property defined in a spacetime point
is smaller. This is investigated for a Schwarzschild metric.
\item A further alternative is to consider models of modified gravity. In the article \cite{Saridakis:2011eq} particle propagation in a
Ho\v{r}ava-Lifshitz modified gravitational background is considered. The authors derive the Dirac equation for a fermion traveling
through such a background. The condition for the existence of nontrivial solutions of the Dirac equation leads to a modified
neutrino dispersion relation. The neutrino velocity can be larger than the speed of light for a special Ho\v{r}ava-Lifshitz scenario.
\item In \cite{Klinkhamer:2011pw} the existence of a sterile neutrino that travels with a superluminal velocity is proposed.
Sterile neutrinos cleverly get around the Cohen--Glashow bound, since they do not couple to the Z boson. An analysis involving
sterile superluminal neutrinos is presented e.g. in \cite{Winter:2011zf}.

\end{itemize}
Furthermore, assuming superluminal neutrino propagation at a certain energy, neutrinos may propagate with a velocity $v\gg c$
at very high energies leading to a different neutrino horizon. In \cite{Lacki:2011wv} bounds from astrophysical observations are
set on $v/c$ for very high neutrino energies.

\section{Extension of the Lorentz group --- neutrinos and a hidden sector}

Let us, for now, take Eq. \eqref{eq:opera-result-neutrino-velocity} as it stands, since the idea of superluminal muon neutrinos
proposed in the current article is purely theoretical and does not rely on the OPERA result. The goal is to describe a
superluminal neutrino species without quantum gravity effects, but by the introduction of new particles coupling to neutrinos.
For the analysis presented as follows the OPERA value can be chosen just as an example (independently of its correctness) in order
to demonstrate the model proposed. That is why we will often refer to Eq.~\eqref{eq:opera-result-neutrino-velocity} in the rest of
the paper.

\subsection{Invariant and maximum velocity}

The foundations of special relativity are the relativity principle and the \textit{constancy} (\textit{invariance}) of the speed
of light. As a result, the Galilei group of classical mechanics is replaced by the Lorentz group, which leads, for example, to
the relativistic law of addition of velocities. The fact that the speed of light is the maximum attainable velocity of all
particles does not directly follow from the Lorentz group, since it only delivers an \textit{invariant} velocity at first. In order
to understand this, three \textit{Gedankenexperiments} will be performed, whose concept was initiated in \cite{Liberati:2001sd}.
\begin{figure*}[t]
\centering
\includegraphics{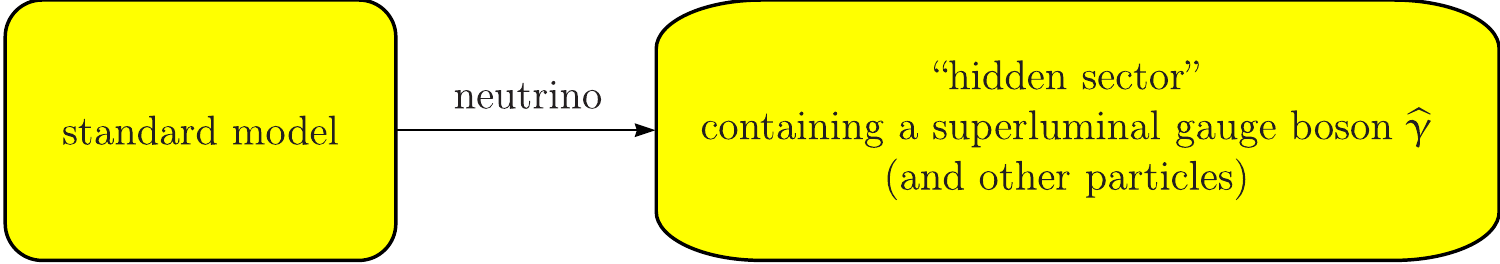}
\caption{Hidden sector that decouples from all the standard model particles except the neutrino. The neutrino is assumed to carry a
charge, which massless hidden sector gauge bosons ${\widehat{\upgamma}}$ couple to. These move with a velocity
${\widehat{c}}$ that is larger than the speed of light ${c}$.}
\label{fig:toy-model-opera-neutrinos-hidden-sector}
\end{figure*}
\begin{itemize}

\item[1)] We consider some hypothetical beings living in a fluid. They are assumed to consist of fluid atoms, which are held together
solely by phonon-mediated forces. The beings do not feel other forces such as electromagnetism or gravitation. Their dynamics is
expected to be governed by an acoustic Lorentz group with the invariant velocity being the velocity of sound $c_s$ in the fluid
\cite{Liberati:2001sd}. We expect the beings to build a rocket, which can be accelerated by phonon emission. From the relativistic
addition of velocities it follows that the maximum attainable velocity of the rocket is given by $c_s$.
\item[2)] Einstein found that the dynamics of particles in our universe is governed by the ``standard'' Lorentz group with the invariant
speed of light $c$. His theory has been heralded or substantiated by various experiments \cite{MichelsonMorley:1887,Kennedy:1932,Ives:1938}.
In the second \textit{Gedankenexperiment} humans build a rocket which is accelerated by a light engine, namely the emission of photons.
In this case, the relativistic addition of velocities leads to $c$ as the upper limit of the rocket velocity.
\item[3)] Now we are ready to discuss the central idea of this article. The basic assumption is that the photon is \textit{not} the gauge
boson which moves with the highest velocity. We adopt neutrinos carrying a new charge $\widehat{q}$ differing from all charges of the
standard model. This charge is to be mediated by a postulated massless gauge boson $\widehat{\gamma}$ moving with a speed $\widehat{c}>c$.
Neutrinos couple to these gauge bosons, which form --- possibly together with other unknown particles --- a hidden sector. The latter does
not interact with any other particle of the standard model, cf. Fig. \ref{fig:toy-model-opera-neutrinos-hidden-sector}. This leads to a
neutrino dynamics which is based on a ``hidden sector Lorentz group'' with an invariant velocity $\widehat{c}$. The third
\textit{Gedankenexperiment} is to build a rocket consisting of neutrinos with an acceleration process working by the emission of gauge
bosons $\widehat{\gamma}$. The limiting velocity of the rocket is then given by $\widehat{c}$, which is larger than the speed of light.

\end{itemize}
\begin{figure*}[t!]
\centering
\includegraphics{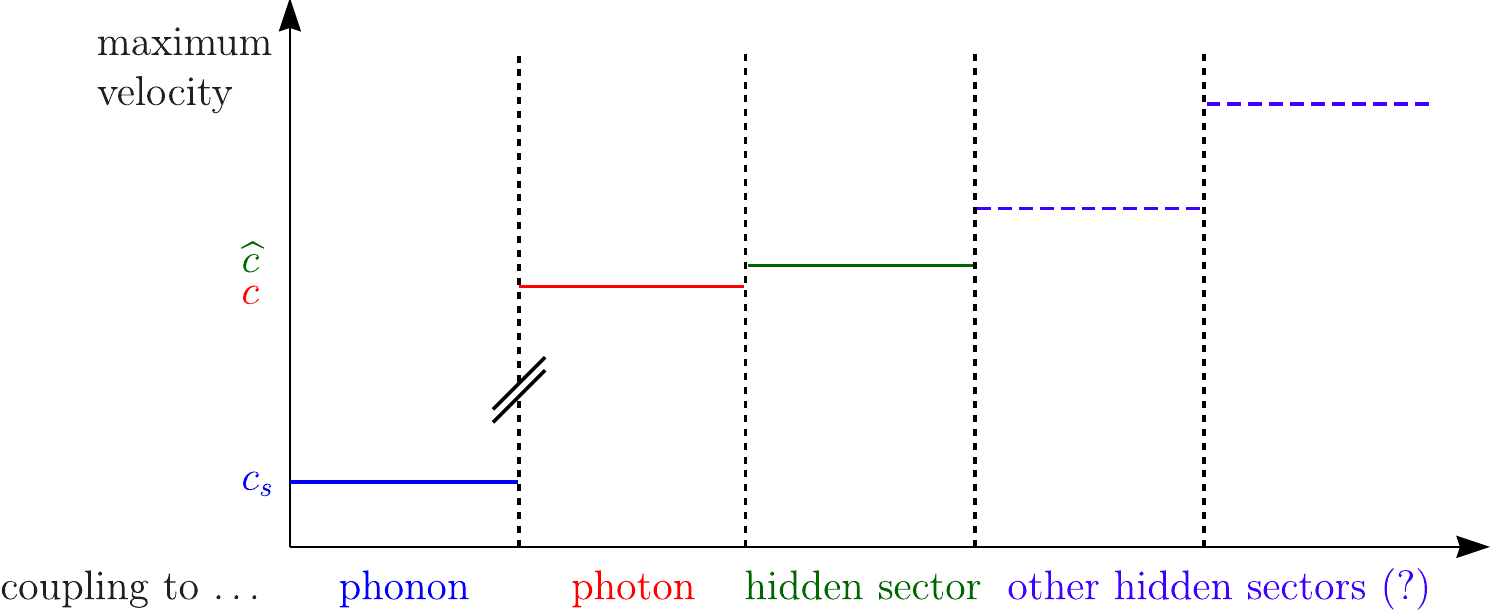}
\caption{Illustration of the situation presented in the previous three \textit{Gedankenexperiments}.
The horizonal axis shows different sectors, each containing a massless boson which transforms under the Lorentz group with a special invariant
velocity. The velocity of this boson sets the maximum attainable velocity for all particles coupling to the corresponding sector. In the first
sector the maximum velocity is given by the speed of sound ${c_s}$ of phonons. Note that the phonon takes a special role here, since
it is not a gauge boson. However, for the very general argument this is not of importance. In the second sector the photon sets the upper limit,
which manifests itself as the speed of light ${c}$, whereas for the hidden sector it is the velocity ${\widehat{c}}$ of the
gauge boson ${\widehat{\gamma}}$. There is the possibility of further sectors whose invariant velocity may also be smaller than
${c}$. If in the latter case a standard model particle couples to such a sector, its kinematics will still be governed by a Lorentz
group with the invariant velocity ${c}$.}
\label{fig:toy-model-opera-neutrinos-invariant-velocities_fig02}
\end{figure*}
The consequences from this deliberation is that different levels of the Lorentz group each with a distinct invariant velocity can be realized
in nature. The coupling constant is assumed to be small enough such that neutrino propagation is not affected too much to violate existing
bounds on the interaction of neutrinos with matter. The size of the coupling is not important for now --- only its existence. The situation
described is depicted in Fig.~\ref{fig:toy-model-opera-neutrinos-invariant-velocities_fig02}. Each new coupling of particles to massless
gauge bosons\footnote{Remark that the phonon is not a gauge boson, but a Goldstone boson (\textit{massless} excitation) resulting from the
spontaneously broken translation symmetry in a solid. This does not matter for the argument, though.} opens a new sector with a maximum
attainable velocity for these particles from left to right.
\subsection{Modified neutrino kinematics and Lagrangian of the hidden sector}
\label{subsec:kinematica-lagrangian}

The gauge boson $\widehat{\upgamma}$ is assumed to couple to neutrino mass eigenstates. This seems a more natural choice than the coupling to flavor
eigenstates, since the hidden sector does not know anything about neutrino flavors. Hence, the neutrino mass eigenstates $\nu_i$ ($i\in \{1,2,3\}$)
obey a kinematics resting upon special relativity, but with the speed of light replaced by the speed $\widehat{c}$ of the hidden sector gauge boson
$\widehat{\gamma}$:
\begin{align}
\label{eq:dispersion-relation-neutrinos-modified}
E_{\upnu_{\scalebox{0.4}{$i$}}}&=\sqrt{\big(m_i\widehat{c}^{\,\,2}\big)^2+(p_{\upnu}\widehat{c}\,)^2}\simeq p_{\upnu}\widehat{c}\Biggl[1+\frac{1}{2}\left(\frac{m_i\widehat{c}}{p_{\upnu}}\right)^2\Biggr] \notag \\
&=\mathcal{G}p_{\upnu} c\Biggl[1+\frac{\mathcal{G}^2}{2}\left(\frac{m_ic}{p_{\upnu}}\right)^2\Biggr]\,,\quad \mathcal{G}=\frac{\widehat{c}}{c}\,.
\end{align}
Here $E_{\upnu_{\scalebox{0.4}{$i$}}}$ is the relativistic neutrino energy of the $i$-th mass eigenstate, $p_{\upnu}$ the neutrino momentum, and $m_i$
its mass.\footnote{We assume that all mass eigenstates propagate with the same momentum $p_{\upnu}$. Whenever we refer to kinematics we stick to the
notation of $m$ with an index (denoting the mass or flavor eigenstate) for the neutrino mass.}
The modified neutrino dispersion relation in Eq.~\eqref{eq:dispersion-relation-neutrinos-modified} is isotropic and $\mathcal{G}$ gives the deformation.
If we suppress the mass eigenstate index $i$ for a moment and interpret neutrinos as matter waves with frequency $\omega$ and three-momentum $k$, we
have to carry out the replacements $E_{\upnu}=\hbar\omega_{\upnu}$ and $p_{\upnu}=\hbar k_{\upnu}$, which leads to:
\begin{equation}
\omega_{\upnu}\simeq \mathcal{G}k_{\upnu}c\Biggl[1+\frac{\mathcal{G}^2}{2}\left(\frac{mc}{\hbar k_{\upnu}}\right)^2\Biggr]\,.
\end{equation}
The front velocity, which corresponds to the velocity of the highest frequency forerunners of a wave, is then given by \cite{Brillouin1960}:
\begin{equation}
v_{\mathrm{fr},\upnu}=\lim_{k_{\upnu}\mapsto \infty} \frac{\omega_{\upnu}}{k_{\upnu}}=\mathcal{G}c\,.
\end{equation}
It equals the signal velocity of a $\delta$-function shaped pulse in configuration space. Hence, any possible distortion of a signal does not play a
role for the front velocity. The case $\mathcal{G}>1$ is related to superluminal, $\mathcal{G}=1$ to luminal, and $\mathcal{G}<1$ to subluminal motion.

If the gauge boson $\widehat{\gamma}$ is assumed to have spin 1 analogously to the photon, at the level of Lagrange densities the ordinary minimal
coupling procedure can be performed with $c$ again replaced by $\widehat{c}$:
\begin{subequations}
\label{eq:lagrangian-hidden-sector}
\begin{equation}
\mathcal{L}^{\text{mass eigenstate }\nu_{\scalebox{0.4}{$i$}}}_{\text{hidden sector}}=\overline{\nu}_{i}\left(\mathrm{i}\gamma^{\mu}\widehat{D}_{\mu}\right)\nu_i
-\left\{\begin{array}{lcl}
\mathcal{L}_{\mathrm{Dirac}}^{(i)} \\
\mathcal{L}_{\mathrm{Maj,1}}^{(i)} \\
\mathcal{L}_{\mathrm{Maj,2}}^{(i)} \\
\end{array}
\right\}\,,
\end{equation}
\begin{align}
\mathcal{L}_{\mathrm{Dirac}}^{(i)}&=\frac{M_D^{(i)}\widehat{c}}{\hbar}(\overline{\nu}_{\scriptscriptstyle{L},i}\nu_{\scriptscriptstyle{R},i}
+\overline{\nu}_{\scriptscriptstyle{R},i}\nu_{\scriptscriptstyle{L},i}) \notag \\
&=\frac{M_D^{(i)}\widehat{c}}{\hbar}\overline{\nu}_{i}\nu_{i}\,,
\end{align}
\begin{equation}
\mathcal{L}_{\mathrm{Maj,1}}^{(i)}=\frac{\widetilde{M}_{1}^{(i)}\widehat{c}}{\hbar}\big(\overline{\nu}^{\,c}_{\scriptscriptstyle{L},i}\nu_{\scriptscriptstyle{L},i}
+\overline{\nu}_{\scriptscriptstyle{L},i}\nu^c_{\scriptscriptstyle{L},i}\big)\,,
\end{equation}
\begin{equation}
\mathcal{L}_{\mathrm{Maj,2}}^{(i)}=\frac{\widetilde{M}_{2}^{(i)}\widehat{c}}{\hbar}\big(\overline{\nu}^{\,c}_{\scriptscriptstyle{R},i}\nu_{\scriptscriptstyle{L},i}
+\overline{\nu}_{\scriptscriptstyle{L},i}\nu^c_{\scriptscriptstyle{R},i}\big)\,,
\end{equation}
\begin{equation}
\widehat{D}_{\mu}=\partial_{\mu}+\mathrm{i}\frac{\widehat{q}}{\hbar}\widehat{A}_{\mu}\,,
\end{equation}
\begin{equation}
\nu_{\scriptscriptstyle{L}}=\frac{\mathds{1}_4-\gamma^5}{2}\nu\,,\quad \nu_{\scriptscriptstyle{R}}=\frac{\mathds{1}_4+\gamma^5}{2}\nu\,,
\end{equation}
\begin{equation}
\label{eq:charge-conjugated-spinor-plus-covariant-derivative}
\nu^c=C\gamma^0\nu^{*}=\mathrm{i}\gamma^2\nu^{*}\,,
\end{equation}
\end{subequations}
where $\{\nu_i,\overline{\nu}_i\equiv \nu_i^{*}\gamma^0\}$ are the neutrino spinor fields describing a specific mass eigenstate and $\gamma^{\mu}$
are the standard Dirac matrices. The covariant derivative $\widehat{D}_{\mu}$ contains the vector field $\widehat{A}_{\mu}$ of the gauge boson
$\widehat{\gamma}$ and the charge $\widehat{q}$, to which $\widehat{\gamma}$ couples. Both a Dirac mass term and two possible choices for Majorana
mass terms \cite{ChengLi:1984} with Dirac mass $M_D$ and Majorana masses $\widetilde{M}_1$, $\widetilde{M}_2$ are given.\footnote{The question,
whether the neutrino is a Dirac or a Majorana particle, has not been answered so far.} Here, $\nu_{\scriptscriptstyle{L}}$ is a left-handed,
$\nu_{\scriptscriptstyle{R}}$ a right-handed neutrino spinor, and $C$ in Eq.~\eqref{eq:charge-conjugated-spinor-plus-covariant-derivative} denotes
the charge conjugation operator.

Note that there is a $\widehat{c}$ in the zeroth component of the partial derivative $\partial_{\mu}$. In the context of the Lorentz-violating
Standard Model Extension \cite{Colladay:1998fq} the nonzero Lorentz-violating coefficients can be found in the left-handed neutrino sector of
$\mathcal{L}_{\mathrm{lepton}}^{\mathit{CPT}\mathrm{-even}}$ in their Eq.~(9).\footnote{Alternatively, the effective Hamiltonian given by Eq.~(14) in
\cite{Kostelecky:2003cr} can be considered.} If we write this term in the mass eigenstate basis and denote the corresponding coefficients by
$(\widetilde{c}_{\scriptscriptstyle{L}})_{ij}$, the $(\widetilde{c}_{\scriptscriptstyle{L}})_{ij}$-matrix is both diagonal in the eigenstate
coefficients $i$, $j$ and diagonal in the Lorentz indices. The latter holds, since the model is isotropic. This leads to
\begin{equation}
(\widetilde{c}_{\scriptscriptstyle{L}})_{\mu\nu,ij}=(\widetilde{c}_{\scriptscriptstyle{L}})_{00}\,\mathrm{diag}\left(1,\frac{1}{3},\frac{1}{3},\frac{1}{3}\right)_{\mu\nu}\delta_{ij}\,.
\end{equation}
The coefficient matrix $(\widetilde{c}_{\scriptscriptstyle{L}})_{\mu\nu}$ is both symmetric and traceless:
\begin{equation}
(\widetilde{c}_{\scriptscriptstyle{L}})_{\mu\nu}=(\widetilde{c}_{\scriptscriptstyle{L}})_{\nu\mu}\,,\quad (\widetilde{c}_{\scriptscriptstyle{L}})^{\mu}_{\phantom{\mu}\mu}=0\,.
\end{equation}
This resembles the \textit{CPT}-even nonbirefringent modified Maxwell theory coefficients $\widetilde{\kappa}_{\mu\nu}$ in the photon sector
\cite{BaileyKostelecky2004,Altschul:2006zz}, which is clear, since both sectors are related by a coordinate transformation --- at least at first
order in the Lorentz-violating coefficients \cite{Altschul:2006zz}.

\subsection{Extension of the toy model to three neutrino flavors}

The neutrino masses $m_i$ are eigenvalues to the mass eigenstates $|\nu_i\rangle$. However, the weak interaction gauge bosons couple to flavor
eigenstates $|\nu_{\alpha}\rangle$ with $\alpha \in\{e,\mu,\tau\}$. The transformation from mass to flavor eigenstates and (vice versa) is governed
by the unitary $(3\times 3)$-PNMS matrix~$U$:
\begin{subequations}
\begin{equation}
|\nu_{\alpha}\rangle=\sum_{i=1,2,3} U_{\alpha,i}|\nu_i\rangle\,,\quad |\nu_i\rangle=\sum_{\alpha=e,\mu,\tau} U_{i,\alpha}^{*}|\nu_{\alpha}\rangle\,.
\end{equation}
When, for simplicity, the \textit{CP}-violating phases are set to zero,\footnote{Furthermore, currently no experimental data
concerning these phases are on hand \cite{Nakamura:2010zzi}.} the matrix $U$ reads
\begin{align}
U&=\begin{pmatrix}
1 & 0 & 0 \\
0 & \phantom{-}\mathrm{c}_{23} & \mathrm{s}_{23} \\
0 & -\mathrm{s}_{23} & \mathrm{c}_{23} \\
\end{pmatrix}\begin{pmatrix}
\phantom{-}\mathrm{c}_{13} & 0 & \mathrm{s}_{13} \\
0 & 1 & 0 \\
-\mathrm{s}_{13} & 0 & \mathrm{c}_{13} \\
\end{pmatrix}
\begin{pmatrix}
\phantom{-}\mathrm{c}_{12} & \mathrm{s}_{12} & 0 \\
-\mathrm{s}_{12} & \mathrm{c}_{12} & 0 \\
0 & 0 & 1 \\
\end{pmatrix} \notag \\
&=\begin{pmatrix}
\mathrm{c}_{12}\mathrm{c}_{13} & \mathrm{s}_{12}\mathrm{c}_{13} & \mathrm{s}_{13} \\
-\mathrm{s}_{12}\mathrm{c}_{23}-\mathrm{c}_{12}\mathrm{s}_{23}\mathrm{s}_{13} & \phantom{-}\mathrm{c}_{12}\mathrm{c}_{23}-\mathrm{s}_{12}\mathrm{s}_{23}\mathrm{s}_{13} & \mathrm{s}_{23}\mathrm{c}_{13} \\
\phantom{-}\mathrm{s}_{12}\mathrm{s}_{23}-\mathrm{c}_{12}\mathrm{c}_{23}\mathrm{s}_{13} & -\mathrm{c}_{12}\mathrm{s}_{23}-\mathrm{s}_{12}\mathrm{c}_{23}\mathrm{s}_{13} & \mathrm{c}_{23}\mathrm{c}_{13} \\
\end{pmatrix}\,,
\end{align}
\end{subequations}
where, for brevity, $\mathrm{s}_{ij}\equiv \sin\theta_{ij}$ and $\mathrm{c}_{ij}\equiv \cos\theta_{ij}$ have been used \cite{Nakamura:2010zzi}. Here,
$\theta_{ij}$ for $(i,j)\in \{(1,2),(1,3),(2,3)\}$ are the neutrino mixing angles. Kinematik measurements of neutrino masses (e.g. for pion decay
and beta decay\footnote{Neutrinoless double beta decay that occurs for Majorana neutrinos leads to a different definition of the ``flavor eigenstate
mass''~\cite{arXiv:0909.2104}.}) lead to ``masses of neutrino flavors,'' which are the weighted average of the neutrino mass eigenvalues
\cite{arXiv:0909.2104}:
\begin{subequations}
\begin{equation}
m_{\upnu_{\scalebox{0.4}{$e$}}}^2\equiv \sum_{i=1,2,3} |U_{e,i}|^2m_i^2\,,
\end{equation}
\begin{equation}
m_{\upnu_{\scalebox{0.4}{$\mu$}}}^2\equiv \sum_{i=1,2,3} |U_{\mu,i}|^2m_i^2\,,
\end{equation}
\begin{equation}
m_{\upnu_{\scalebox{0.4}{$\tau$}}}^2\equiv \sum_{i=1,2,3} |U_{\tau,i}|^2m_i^2\,.
\end{equation}
\end{subequations}
Since in Eq. \eqref{eq:dispersion-relation-neutrinos-modified} the neutrino mass eigenvalues $m_i$ are multiplied by $\widehat{c}_i^{\,\,2}$, the
maximum velocity of each neutrino flavor will be defined in the following way:
\begin{subequations}
\begin{equation}
\widehat{c}_{\upnu_{\scalebox{0.4}{$e$}}}^{\,\,4}\equiv \sum_{i=1,2,3} |U_{e,i}|^2\,\widehat{c}_i^{\,\,4}\,,
\end{equation}
\begin{equation}
\widehat{c}_{\upnu_{\scalebox{0.4}{$\mu$}}}^{\,\,4}\equiv \sum_{i=1,2,3} |U_{\mu,i}|^2\,\widehat{c}_i^{\,\,4}\,,
\end{equation}
\begin{equation}
\widehat{c}_{\upnu_{\scalebox{0.4}{$\tau$}}}^{\,\,4}\equiv \sum_{i=1,2,3} |U_{\tau,i}|^2\,\widehat{c}_i^{\,\,4}\,.
\end{equation}
\end{subequations}
If we assume
\begin{subequations}
\begin{equation}
\delta\widehat{c}_i/\widehat{c}_i\ll 1\,,\quad \delta\widehat{c}_i\equiv \widehat{c}_i-c\,,
\end{equation}
\begin{equation}
\delta \widehat{c}_{\upnu_{\scalebox{0.4}{$\alpha$}}}/\widehat{c}_{\upnu_{\scalebox{0.4}{$\alpha$}}}\ll 1\,,\quad \delta\widehat{c}_{\upnu_{\scalebox{0.4}{$\alpha$}}}\equiv \widehat{c}_{\upnu_{\scalebox{0.4}{$\alpha$}}}-c\,,
\end{equation}
\end{subequations}
it is sufficient to linearize the equations above:
\begin{align}
c^4+4c^3\delta\widehat{c}_{\upnu_{\scalebox{0.4}{$\alpha$}}} &\simeq \sum_{i=1,2,3} |U_{\alpha,i}|^2(c^4+4c^3\delta\widehat{c}_i) \notag \\
&=c^4+4c^3\sum_{i=1,2,3} |U_{\alpha,i}|^2\delta\widehat{c}_i\,.
\end{align}
Here, $\alpha\in \{e,\mu,\tau\}$. From the latter equation follows the simplified result
\begin{subequations}
\label{eq:connection-between-calpha-and-ci}
\begin{equation}
\delta\widehat{c}_{\upnu_{\scalebox{0.4}{$\alpha$}}}\simeq \sum_{i=1,2,3} |U_{\alpha,i}|^2\delta\widehat{c}_i\,,
\end{equation}
\begin{equation}
\Rightarrow \widehat{c}_{\upnu_{\scalebox{0.4}{$\alpha$}}}\simeq \sum_{i=1,2,3} |U_{\alpha,i}|^2\,\widehat{c}_i\,.
\end{equation}
\end{subequations}
In the rest of the article the neutrino velocities will be approximated by $\widehat{c}_i$ for the mass eigenstates and by
$\widehat{c}_{\upnu_{\scriptscriptstyle{\alpha}}}$ for the flavor eigenstates, since neutrino masses are assumed to be much smaller than
neutrino energies.

\subsubsection{The Supernova 1987 bound}

The point of extending the toy model to all active neutrino flavors, is to account for the Supernova 1987 (SN 1987) bound on electron antineutrinos\footnote{
An antineutrino is assumed to travel with the same velocity as the corresponding neutrino. This makes sense, since the model presented corresponds
to a \textit{CPT}-even term of the Lorentz-violating Standard Model Extension. See the end of Sec.~\ref{subsec:kinematica-lagrangian} for a brief discussion
concerning this issue.} \cite{Hirata:1987hu,Longo:1987ub}:
\begin{equation}
\label{eq:sn1987-bound-electron-neutrinos}
\left|\frac{(v_{\overline{\upnu}_{\scriptscriptstyle{e}}}|_{E_{\overline{\upnu}_{\scalebox{0.4}{$e$}}}\in [7.5,36]\,\mathrm{MeV}})-c}{c}\right|\lesssim 2\cdot 10^{-9}\,.
\end{equation}
An electron neutrino produced by a weak interaction process is a mixture of neutrino mass eigenstates according to the relation
\begin{equation}
|\nu_e\rangle=\sum_{i=1,2,3} U_{e,i}|\nu_i\rangle\,.
\end{equation}
After the neutrino has traveled through space and reaches a distance $L$ from the origin of its production it holds
\begin{align}
|\nu_e(L)\rangle&\approx \sum_{i=1,2,3} U_{e,i}\exp\left(-\mathrm{i}\frac{m_i^2}{2E}L\right)|\nu_i\rangle \notag \\
&=\sum_{\beta=e,\mu,\tau}\left[\sum_{i=1,2,3} U_{e,i}\exp\left(-\mathrm{i}\frac{m_i^2}{2E}L\right)U_{i,\beta}^{*}\right]|\nu_{\beta}\rangle\,,
\end{align}
where $m_i\ll E$ has been assumed \cite{Nakamura:2010zzi}. Hence, the initial electron neutrino state then corresponds to a mixture of all flavor
eigenstates. However, the initial composition of mass eigenstates remains the same, because quantum mechanically the statement
\begin{align}
|\langle \nu_j|\nu_e(L)\rangle|^2&=\Big|\sum_{i=1,2,3} U_{e,i}\exp\left(-\mathrm{i}\frac{m_i^2}{2E}L\right)\delta_{ij}\Big|^2 \notag \\
&=\Big|U_{e,j}\exp\left(-\mathrm{i}\frac{m_j^2}{2E}L\right)\Big|^2=|U_{e,j}|^2\,,
\end{align}
is valid. As a result of that, also the velocity of the neutrino does not change during its propagation, since it is determined by the initial
composition of mass eigenstates. The antineutrinos coming from the supernova were detected as electron antineutrinos on Earth. For this reason
the bound of Eq.~\eqref{eq:sn1987-bound-electron-neutrinos} will be considered as a bound on the velocity of electron neutrinos --- regardless
of whether their flavor was different on their way to Earth.

We assume three distinct hidden sectors each with its own gauge boson $\widehat{\upgamma}_{i}$, where $\widehat{\upgamma}_1$ only couples to the
first mass eigenstate, $\widehat{\upgamma}_2$ to the second, and $\widehat{\upgamma}_3$ to the third, via the respective charge $\widehat{q}_i$
for $i=1$, 2, and 3, respectively. If any sector obeys a different invariant velocity $\widehat{c}_1\neq \widehat{c}_2\neq \widehat{c}_3$, the
constraint of Eq.~\eqref{eq:sn1987-bound-electron-neutrinos} does not necessarily contradict a deviation from the speed of light of the order of
$10^{-5}$ for one single neutrino flavor. This will be shown as follows.

The current experimental values or bounds for the three neutrino mixing angles $\theta_{12}$, $\theta_{23}$, and $\theta_{13}$ are
\cite{Nakamura:2010zzi}:
\begin{subequations}
\begin{equation}
\sin^2(2\theta_{12})=0.87\pm 0.03\,,
\end{equation}
\begin{equation}
\sin^2(2\theta_{23})>0.92\,,
\end{equation}
\begin{equation}
\sin^2(2\theta_{13})<0.15\,.
\end{equation}
\end{subequations}
With the lower bound on $\theta_{23}$ and the upper bound on $\theta_{13}$ we obtain the PNMS matrix
\begin{equation}
U\approx \begin{pmatrix}
0.81 & 0.55 & 0.20 \\
-0.55 & 0.59 & 0.59 \\
0.21 & -0.58 & 0.79 \\
\end{pmatrix}\,.
\end{equation}
Current experimental data imply that neutrinos are almost massless. Concretely, this means $m_{\upnu_{\scalebox{0.4}{$e,\mu,\tau$}}}\lesssim 1\,\mathrm{eV}/c^2$
from neutrino oscillation data \cite{Nakamura:2010zzi} and\footnote{Note that under the assumptions taken, the unit $\mathrm{eV}/c^2$ should be replaced by
$\mathrm{eV}/\widehat{c}^{\,2}$, as well. But since the mass values given have been obtained in the context of special relativity, where the speed of light $c$
is the invariant velocity, we keep $c$.}
\begin{equation}
\sum_{f=e,\mu,\tau} m_{\upnu_{\scalebox{0.4}{$f$}}}<0.67\,\mathrm{eV}/c^2\, (95\%\,\mathrm{CL})\,,
\end{equation}
which is obtained from WMAP observations \cite{Komatsu:2008hk}. Therefore, an approximate value of $\widehat{c}_{\upnu_{\scalebox{0.4}{$\mu$}}}$ directly follows
from Eq.~\eqref{eq:opera-result-neutrino-velocity}:
\begin{equation}
\widehat{c}_{\upnu_{\scalebox{0.4}{$\mu$}}}=\mathcal{G}_{\mu}c\,,\quad \mathcal{G}_{\mu}\simeq 1 + 2.37\cdot 10^{-5}\,.
\end{equation}
Assuming $\widehat{c}_3=c$ and $\widehat{c}_{\upnu_{\scalebox{0.4}{$e$}}}=c$ we obtain:\footnote{The latter choice is reasonable, since $|(v_{\upnu_{\scriptscriptstyle{e}}}-c)/(v_{\upnu_{\scriptscriptstyle{\mu}}}-c)|_{\mathrm{exp}}\lesssim  8.4\cdot 10^{-5}$ according to
Eqs.~\eqref{eq:opera-result-neutrino-velocity} and \eqref{eq:sn1987-bound-electron-neutrinos}. We keep in mind that the constraints on
$v_{\upnu_{\scriptscriptstyle{e}}}$ and $v_{\upnu_{\scriptscriptstyle{\mu}}}$ were obtained at different neutrino energies, but this does
not play a role in our model, though.}
\begin{equation}
\frac{\widehat{c}_1-c}{c}\approx -5.30\cdot 10^{-5}\,,\quad \frac{\widehat{c}_2-c}{c}\approx 1.13\cdot 10^{-4}\,.
\end{equation}
Hence, Eq.~\eqref{eq:opera-result-neutrino-velocity} for muon neutrinos and the SN 1987 bound for electron neutrinos do not clash, if the velocity
of the first mass eigenstate is a little bit lower than $c$ and if the second moves faster than $c$. Since the first eigenstate propagates slower
than $c$, it need not necessarily couple to any hidden sector. In contrast, the second mass eigenstate has to couple to a $\widehat{\upgamma}_2$
traveling faster than light.

\subsection{Challenges of the model and introduction of sterile neutrinos}

The argument of \cite{Cohen:2011hx} resulting in the rapid loss of the neutrino energy by electron positron emission relies on fundamental
principles: four-momentum conservation and the coupling of the neutrino sector to the Z boson. Models for superluminal neutrino propagation have
to compete with the very general result mentioned, and this is also the case for the toy model presented here.
\begin{itemize}

\item[1)] We could assume the energy loss of muon neutrinos to be compensated by a Compton scattering type process, where gauge bosons
$\widehat{\upgamma}$ scatter with muon neutrinos. However, this argument leads to additional problems. First of all, the free parameters of the
model (e.g. the charge $\widehat{q}$ or the initial energy of a $\widehat{\upgamma}$ boson) have to be chosen such that this compensation is
possible, which requires extreme finetuning. If the momentum distribution of $\widehat{\gamma}$ is homogeneous and isotropic, the average energy
transfer to the neutrino will be zero. In principle, the distribution may be anisotropic, but then neutrinos might be deflected on their way from
CERN to the Gran Sasso underground laboratory.
\item[2)] An alternative proposal is that a neutrino itself is part of the hidden sector making it to some kind of superluminal, \emph{sterile}
neutrino $\upnu_s$. Then the neutrino does not couple to the Z boson, rendering the process $\upnu_s\rightarrow \upnu_s\mathrm{e^+}\mathrm{e^-}$
forbidden. The sterile neutrino may mix with the active neutrino species leading to superluminal propagation of at least some of the standard model
neutrino flavors. This idea has already been suggested in other publications, see e.g. \cite{Klinkhamer:2011pw,Klinkhamer:2011js} and references
therein. Reference \cite{Marfatia:2011bw} states that sterile neutrino models may be in conflict with the atmospheric neutrino data measured at
Super-Kamiokande. However, the models considered in the latter article only involve one sterile neutrino and one single mixing angle with this
neutrino. Conclusions for models with more mixing angles have not been obtained.

\end{itemize}
\begin{figure*}[t!]
\centering
\subfigure[\,\,\,$\theta_{14}=\pi/3$]{\includegraphics[scale=0.45]{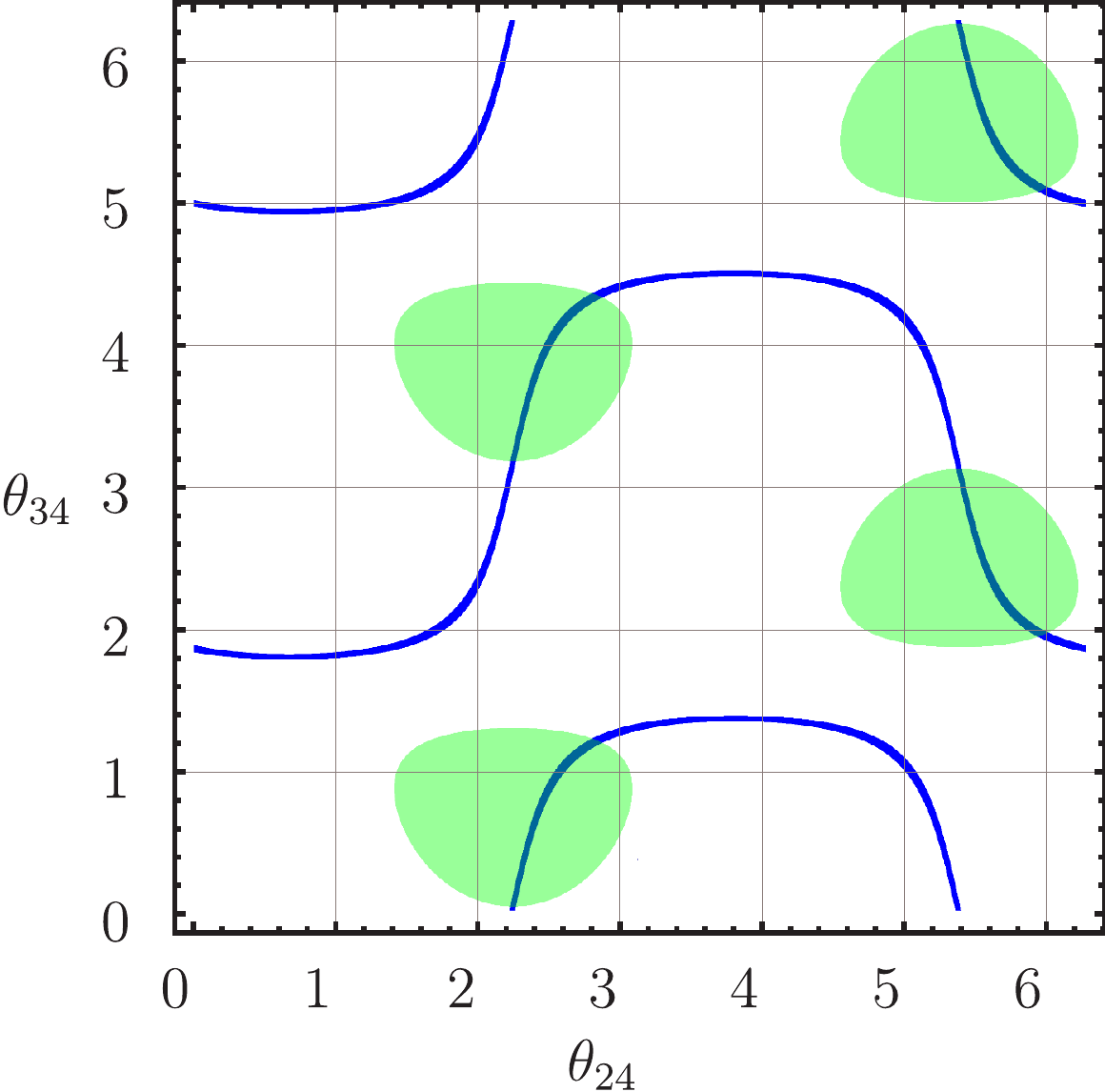}}
\hspace{0.5cm}\subfigure[\,\,\,$\theta_{24}=\pi/5$]{\includegraphics[scale=0.45]{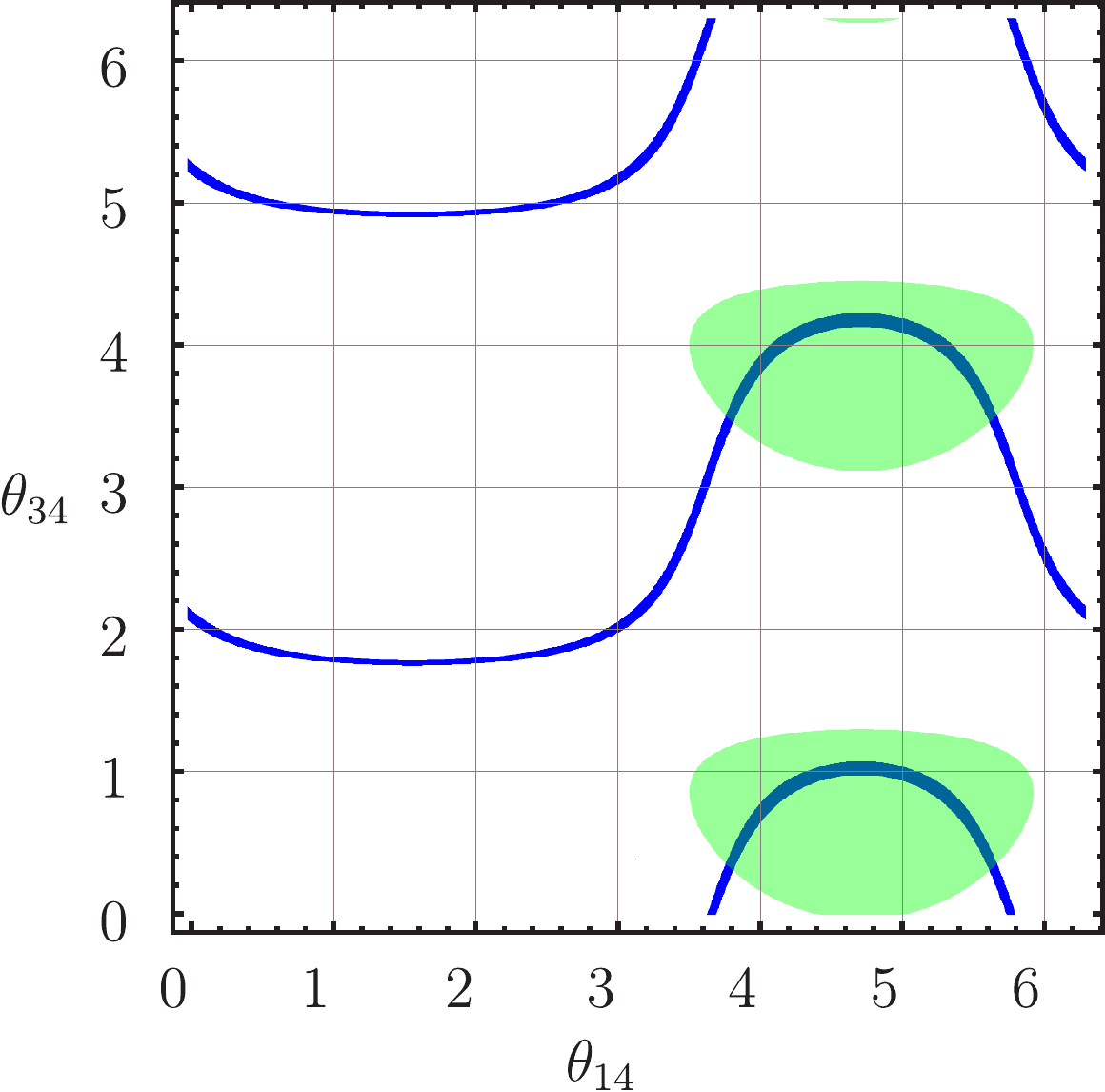}}
\hspace{0.5cm}\subfigure[\,\,\,$\theta_{34}=\pi/5$]{\includegraphics[scale=0.45]{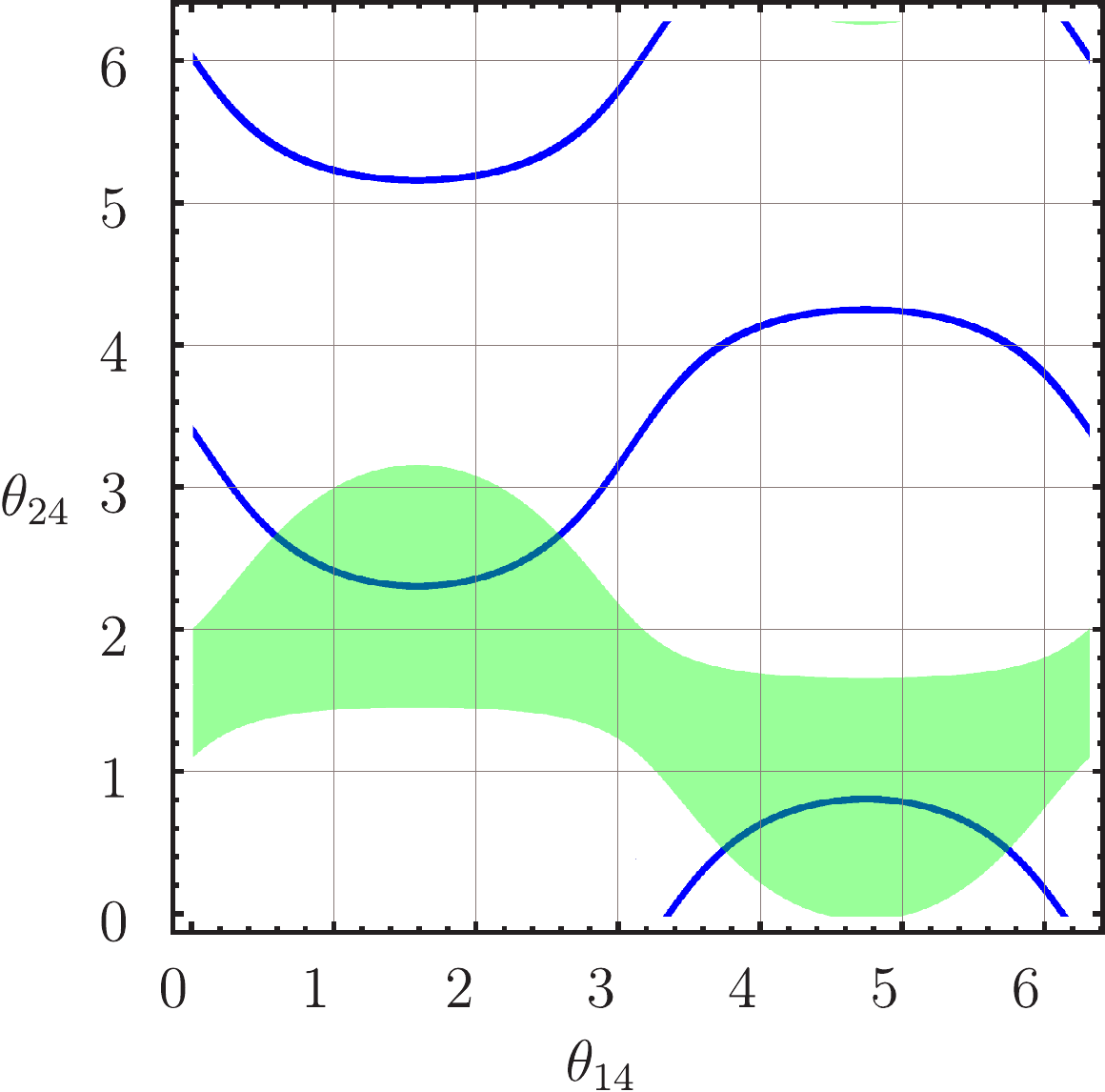}}
\caption{Each panel shows the plane of a different pair of sterile neutrino mixing angles, where the remaining angle is set to the special value given
below the corresponding panel. Regions of the electron and the muon neutrino velocity are shown, where ${\widehat{c}_1=\widehat{c}_2=\widehat{c}_3=c}$,
and ${\widehat{c}_4=(1+3\cdot 10^{-5})c}$ is the common choice. The blue areas depict the region ${(\widehat{c}_{\upnu_e}/c - 1)< 10^{-8}}$
and the green areas show the region ${(\widehat{c}_{\upnu_\mu}/c - 1)> (2.37-0.32-0.24)\cdot 10^{-5}=1.81\cdot 10^{-5}}$. The condition
${(\widehat{c}_{\upnu_\mu}/c - 1) <(2.37+0.32+0.34)\cdot 10^{-5}=3.03\cdot 10^{-5}}$ is fulfilled for all possible mixing angles in each panel,
so is ${(\widehat{c}_{\upnu_e}/c - 1)> -10^{-8}}$.}
\label{fig:toy-model-opera-neutrinos-regionplot}
\end{figure*}
According to the second item of the list above, we extend the toy model by $N_s$ sterile neutrino mass eigenstates. Then the transformation between
the $3+N_s$ flavor and mass eigenstates is governed by a unitary~$(3+N_s)\times(3+N_s)$-matrix $U^s$:
\begin{equation}
|\nu_{\alpha}\rangle=\sum_{i=1}^{3+N_s} U^s_{\alpha,i}|\nu_i\rangle\,,
\end{equation}
Following Eq. \eqref{eq:connection-between-calpha-and-ci} we can write:
\begin{equation}
\widehat{c}_{\upnu_{\scalebox{0.4}{$\alpha$}}}\simeq \sum_{i=1}^{3+N_s} |U^s_{\alpha,i}|^2\,\widehat{c}_i\,.
\end{equation}
For our toy model we consider the simplest case with one single sterile neutrino, hence $N_s=1$. In principle, this sterile neutrino mass eigenstate
$\nu_4$ mixes with the active neutrino mass eigenstates $\nu_i$ for $i\in \{1,2,3\}$. This mixing can be described by introducing three additional
mixing angles $\theta_{14}$, $\theta_{24}$, and $\theta_{34}$. The corresponding $(4\times 4)$-mixing matrix $U^s$ can then be constructed from $U$
as follows:\footnote{with all \textit{CP}-violating phases set to zero}
\begin{align}
U^s&\equiv \begin{pmatrix}
U & \mathbf{0} \\
\mathbf{0}^{\intercal} & 1 \\
\end{pmatrix} \begin{pmatrix}
\phantom{-}\mathrm{c}_{14} & 0 & 0 & \mathrm{s}_{14} \\
0 & 1 & 0 & 0 \\
0 & 0 & 1 & 0 \\
-\mathrm{s}_{14} & 0 & 0 & \mathrm{c}_{14} \\
\end{pmatrix} \notag \\
&\quad\,\times \begin{pmatrix}
1 & 0 & 0 & 0 \\
0 & \phantom{-}\mathrm{c}_{24} & 0 & \mathrm{s}_{24} \\
0 & 0 & 1 & 0 \\
0 & -\mathrm{s}_{24} & 0 & \mathrm{c}_{24} \\
\end{pmatrix} \begin{pmatrix}
1 & 0 & 0 & 0 \\
0 & 1 & 0 & 0 \\
0 & 0 & \phantom{-}\mathrm{c}_{34} & \mathrm{s}_{34} \\
0 & 0 & -\mathrm{s}_{34} & \mathrm{c}_{34} \\
\end{pmatrix}\,,
\end{align}
where $\mathbf{0}=(0,0,0)$. In what follows, we examine a subspace of the free toy model parameters, which is seven-dimensional. It is spanned
by the three sterile neutrino mixing angles $\theta_{14}$, $\theta_{24}$, $\theta_{34}$ and by the invariant velocities of the neutrino mass eigenstates
$\widehat{c}_1$, $\widehat{c}_2$, $\widehat{c}_3$, $\widehat{c}_4$. Since this phase space is that large, it will be reduced by the special choice below.
We assume that the invariant velocities of the three standard neutrino mass eigenstates correspond to the speed of light, which means
$\widehat{c}_1=\widehat{c}_2=\widehat{c}_3=c$. The single sterile neutrino is assumed to travel with superluminal speed: we therefore set
$\widehat{c}_4=(1+3\cdot 10^{-5})c$.

As a result, only the sterile mixing angles remain as free parameters. In Fig.~\ref{fig:toy-model-opera-neutrinos-regionplot} three cases are considered,
where in each one of these angles is fixed: $\theta_{14}=\pi/3$, $\theta_{24}=\pi/5$, and $\theta_{34}=\pi/5$. We would like to explore, whether in each
case the remaining two angles can be chosen such that the electron anti-neutrino velocity respects the SN 1987 bound of
Eq.~\eqref{eq:sn1987-bound-electron-neutrinos} and that the muon neutrino velocity lies in the error band of Eq.~\eqref{eq:opera-result-neutrino-velocity}.
In all plots overlapping regions are small, but they exist. At least one of the three mixing angles has to be rather large. A special possibility is
$\theta_{14}=5\pi/4$, $\theta_{24}=\pi/5$, and $\theta_{34}=\pi/5$ that becomes evident from the third panel. With these values we obtain the following
results for the velocities of the three active neutrino flavors:
\begin{subequations}
\begin{equation}
\frac{\widehat{c}_{\upnu_{\scriptscriptstyle{e}}}-c}{c}\approx 9.07\cdot 10^{-10}\,,
\end{equation}
\begin{equation}
\frac{\widehat{c}_{\upnu_{\scriptscriptstyle{\mu}}}-c}{c}\approx 2.33\cdot 10^{-5}\,,
\end{equation}
\begin{equation}
\frac{\widehat{c}_{\upnu_{\scriptscriptstyle{\tau}}}-c}{c}\approx 2.31\cdot 10^{-7}\,.
\end{equation}
\end{subequations}
Whether there exists a choice of angles that does not contradict existing atmospheric neutrino data --- as was proposed in \cite{Marfatia:2011bw} --- will
not be examined here.

To summarize, within the toy model presented a superluminal sterile neutrino mass eigenstate can be introduced, such that the electron neutrino respects the
SN 1987 bound and the muon neutrino travels with the superluminal velocity that is given by Eq.~\eqref{eq:opera-result-neutrino-velocity}. For the parameters
chosen above the tau neutrino is then slightly superluminal, as well.

\section{Conclusions}

In this article a concept accounting for superluminal muon neutrinos was presented. It is based on a multiple Lorentz group structure. The dynamics of the
neutrino is assumed to obey the Lorentz group with an invariant velocity that is larger than the speed of light. This will be possible, if the neutrino
couples to a hidden sector of massless gauge bosons that move faster than photons. Then the neutrino field transforms under the Lorentz group with an invariant
velocity which corresponds to the velocity of these gauge bosons.

If an experiment measures a deviation of the neutrino velocity that is much larger than the speed of light, this will be very difficult to understand in the
context of physics at the Planck scale. The idea presented here leads, in principle, to a modified dispersion relation of the neutrino, as well. However, the
framework is not quantum gravity, but special relativity and field theory with an invariant velocity imposed that differs from the speed of light.

First of all, every physical model describing superluminal muon neutrinos has to compete with the SN 1987 bound. This is a minor difficulty, since the toy model
presented here can be altered such that electron neutrinos behave differently compared to muon neutrinos. More severe is the Cohen--Glashow constraint that is
based on fundamental principles of present-day physics, which are difficult to circumvent. Honestly, the latter is also a severe problem for the current model,
unless the superluminal neutrino itself is part of a hidden sector, hence sterile.

Besides that, the toy model makes the following predictions that can be verified or falsified by experiment:
\begin{itemize}

\item If the muon neutrino moves with a superluminal velocity, its velocity is isotropic and does not depend on the neutrino energy (besides any mass-related
dependence).
\item It is \emph{not} a local effect, i.e. muon neutrinos move with a superluminal velocity in interstellar space, as well.

\end{itemize}

To find a physical theory both explaining superluminal neutrinos without bothering already established data and facts about the neutrino sector is a great
challenge for model builders.

\section*{Acknowledgments}

It is a pleasure to thank S.~Thambyahpillai (KIT) and A.~Crivellin (University of Bern) for reading an early draft of the paper and for useful suggestions.
Furthermore, the author thanks V.~A.~Kosteleck\'{y} (Indiana University, Bloomington) for helpful discussions.

The author acknowledges support by \textit{Deutsche Forschungsgemeinschaft} and Open Access Publishing Fund of Karlsruhe Institute of Technology.

\end{document}